**Spatial Small Worlds: New Geographic Patterns for an Information Economy** [1]

Sean P. Gorman and Rajendra Kulkarni

*School of Public Policy, George Mason University, Finley Building, 4400 University Drive, Fairfax, VA 22030, USA*

Networks are structures that pervade many natural and man-made phenomena. Recent findings have characterized many networks as not random chaotic structures but as efficient complex formations. Current research has examined complex networks as largely a non-spatial phenomenon. Location, distance, and geography, though, are all vital aspects of a wide variety of networks. This paper will examine the United States' portion of the Internet's infrastructure as a complex networks and what role distance and geography play in its formation. From these findings implications will be drawn on the economic, political, national security and technological impacts of network formation and evolution in an information economy.

## Introduction

Networks are ubiquitous in the everyday world, some obvious – the network of roads driven on, the fiber networks connecting computers – and some not so obvious – the economic networks that enable globalization, molecular networks that keep human bodies functioning. Surprisingly these networks have many things in common, and understanding the complex and evolving nature of them has garnered an increasing amount of interest. All networks share a common construct of nodes connected together by links. The very simple concept of one location connecting to another quickly becomes an extremely complex phenomenon as the number of nodes and connections increase. Figuring out how these simple concepts evolve into incredibly complex and dynamic networks has produced a flurry of work in physics, computer science, molecular biology, sociology, and many other fields. The one aspect of networks commonly overlooked in this new field of inquiry is the geography of networks. The vast majority of research on complex networks revolves around abstract networks where geographic location is not considered.

[1] The authors would like to thank Critical Infrastructure Protection Project at the George Mason School of Law for their funding of the research. The authors would also like to thank Dr. Ed Malecki and Anthony Townsend for loaning from their respective National Science Foundation grants BCS-9911222 and SBR-9817778. Also many thanks are in order to Martin Dodge and Dr. Laurie Schintler for their help with the mathematical and technical aspects of network theory. Finally the authors would like to thank three anonymous referees for their many constructive ideas and direction.

Many have considered the dearth of attention to location and geography to result because they are too complicated to fit into current network models (Batty 2001). Some initial work by Yook, Jeong, and Barabasi (2001) has examined the role of linear distance in complex networks, but a pure spatial or geographical analysis of complex networks has not been completed. In an attempt to fill the gap in analysis this paper examines the United States' Internet infrastructure as a complex network. This analysis has two steps. First, the USA's Internet infrastructure is analyzed under the framework of existing complex network models developed using statistical physics. Second, a spatial model of complex networks is developed to explain the geographic structure of the USA's Internet infrastructure. Before presenting analysis an overview of complex networks and their study is presented. Lastly the paper concludes with a discussion of what impacts spatial small worlds may have on geography, economics, security, and policy.

**A Review of Small Worlds and Complex Network Research**

The mathematical study of networks commonly falls under graph theory. Graph theory has been used to model a wide array of networks for empirical analysis, including transportation, communications, and neural networks. Sometimes networks are less apparent, as with economics where companies are vertices and transactions between them are edges, or social networks where people are nodes and acquaintance is the edge (Arthur 1999, Wasserman and Faust 1994, Hayes 2000a, p.10). Early graph theory analysis was confined to relatively small networks with a computationally manageable number of edges and vertices. This work included many applications from geographical analysis, especially with transportation networks. Kansky (1963), Garrison (1968), Haggett and Chorley (1969) as well as many others used the graph theoretical implications of transportation networks to help explain aspects of regional and national economies. There was also attention specifically given to the small world problem outlined by Milgram (1977) - how many steps does it take to link any two people, selected at random? - specifically in the context of geography. Stoneham (1977) investigated the spatial aspects of the small world problem, examining, "the general distribution of steps with parameter changes; channeling effects; the sensitivity of the overall structure to disconnection; and ghettoisation of an area" (p.185).

During roughly the same period, Erdös and Renyi (1960) were doing theoretical work focused on large complex graphs. Erdös and Renyi (ER) endeavored to use "probabilistic methods" to solve problems in graph theory, where a large number of nodes were involved (Albert and Barabasi 2002, p. 54). Under this assumption they modeled large graphs utilizing algorithms where $n$ nodes were randomly connected according to probability *P*, and found that when vertices were connected in this fashion they followed a Poisson distribution (Albert and Barabasi 2002, p 49). A more thorough review of random graphs can be found in the survey work of Bollobás (1985). Following ER's findings their random models of network formation were widely used in several disciplines researching networks, the most topical to this research being Internet topology generators (Radoslavov et al 2000).

The absence of detailed topological data for complex networks left random network models as the most widely used method of network simulation (Barabasi 2001). As computing power increased and real world network data began to become available and several empirical findings emerged. Three network characteristics frequently resulted from analysis of complex networks:

1. Short average path length

2. High level of clustering

3. Power law and exponential degree distributions

(Albert and Barabasi 2002, p. 48-49)

Short average path length indicates that the distance between any two nodes on the network is short; they can be reached in a few number of hops along edges. Clustering occurs when nodes locate topologically close to each other in cliques that are well connected to each other. Lastly, the frequency distributions of node density, called degrees, often follow power laws.

Watts and Strogatz (WS) (1998) formalized this concept of clustering for complex networks. Using several large data sets, they found that the real-world networks studied were not entirely random but instead displayed significant clustering at the local level. Further, that local clusters linked across the graph to each other forming "small worlds". To model this effect Watts and Strogatz (1998) took a regular lattice where all neighbors are connected to their two nearest neighbors and randomly rewired nodes in the lattice. These short cuts across the graph to different clusters of vertices introduced a level of efficiency[2] not predicted in the ER model. The distribution was not Poisson as with the ER model, but was bounded and decayed exponentially for large sets of vertices (Watts and Strogatz 1998). Watts (2003) has extended this work recently to cover topics ranging from, "epidemics of disease to outbreaks of market madness, from people searching for information to firms surviving crisis and change, from the structure of personal relationships to the technological and social choices of entire societies (p.1)." The work by WS was not the first, though, to investigate the effects of rewiring:

> …Fan R. K. Chung, in collaborations with Michael R. Garey of AT&T Laboratories and Béla Bolobás of the University of Memphis, studied various ways of adding edges to cyclic graphs. They found cases where the diameter is proportional to log $n$ (Hayes 2000b, p.106).

The finding of WS spurred a flurry of work into understanding the attributes of complex networks and new findings and discoveries quickly followed. Two parallel studies by

---

[2] Efficiency in this case refers to the network characteristic of a large number of nodes having a low diameter.

Albert, Jeong, and Barabasi (1999) of Notre Dame and Huberman and Adamic (1999) at Xerox Parc found that when one looks at the World Wide Web as a graph (web pages are vertices and hyperlinks connecting them are edges) it followed not a Poisson or exponential distribution, but a power law distribution.

A power law is a significantly different finding from either the expected exponential or Poisson distribution. In a power law distribution there is an abundance of nodes with only a few links, and a small but significant minority that have very large number of links (Barabasi 2002). It should be noted that this is distinctly different from both the ER and WS model; the probability of finding a highly connected vertex in the ER and WS model decreases exponentially, thus, "vertices with high connectivity are practically absent"[3] (Barabasi and Albert 1999, p.510). The reason, according to Barabasi and Albert (1999), was that their model added another perspective to complex networks, incorporating network growth; the number of nodes does not stay constant as in the WS and ER model. The BA models added growth over time and the idea that new vertices attach preferentially to already well-connected vertices in the network

Barabasi and Albert (1999) formalized this idea in "Emergence of Scaling in Random Networks". They stated that in a complex network like the World Wide Web the probability $P(k)$ that a vertex in the network interacts with $k$ other vertices decays as a power law following $P(k) \sim k^{-g}$ where the power law exponent is equal to three (see

---

[3] Barabasi and Albert's definition of high connectivity is relative to the number of nodes in the network, and in this context simply means a large proportion on the total connections in the network. The odds of a node having a large proportion on connections in a network are small enough that they are likely to be "practically absent".

Figure 1 for a graphic representation of the function). When studying real world scale free networks empirical results have ranged from 2.1 to 4 (Barabasi and Albert 1999). While the model set up by Barabasi and Albert produces an exponent of three, they demonstrate how the model can be altered to produce results other than three for different network conditions. The Barabasi-Albert (BA) model is based on three mechanisms that drive the evolution of graph structures over time to produce power law relationships:

1. Incremental growth – Incremental growth follows from the observation that most networks develop over time by adding new nodes and new links to existing graph structure.

2. Preferential connectivity – Preferential connectivity expresses the frequently encountered phenomenon that there is higher probability for a new or existing node to connect or reconnect to a node that already has a large number of links (i.e. high vertex degree) than there is to (re)connect to a low degree vertex.

3. Re-wiring – Re-wiring allows for some additional flexibility in the formation of networks by removing links connected to certain nodes and replacing them with new links in a way that effectively amounts to a local type of re-shuffling connection based on preferential attachment.

(Chen et al 2001, p.5)

The difference between the random model of Erdös and Renyi and the model described by Barabasi and Albert becomes clearer when seen in a visual representation. Figure 2 illustrates the structural difference between a random ER network model and a scale free network model.

> The high level of clustering and super-connected node is evident on the node diagram. For the model in Figure 3, more than 60% of nodes (green) can be reached from the five most connected nodes (red) compared with only 27% in the random network. This demonstrates the key role that hubs play in the scale-free network. Both networks contain the 130 nodes and 430 links.
>
> (Barabasi 2001, p.1)

This leaves the rather fuzzy question of what is a small world and what is a scale free network. As stated earlier Albert and Barabasi (2002) see small worlds and scale free networks as explanations for two different phenomena occurring in complex networks. The WS small world model explains clustering and the scale free model explains power law degree distributions (Albert and Barabasi 2002, p.49). Their have, though, been other opinions on how small world and scale free networks should be classified, Amaral et al (2000) posits that scale free networks are a sub class of small world networks. Further, that there are three classes of small world networks:

1. Scale-free networks, characterized by a vertex connectivity distribution that decays as a power law.
2. Broad-scale networks, characterized by a connectivity distribution that has a power law regime followed by a sharp cutoff.

3. Single-scale networks, characterized by a connectivity distribution with a fast decaying tail.

(Amaral et al 2000, p.11149)

An exact delineation of where small world and scale free networks diverge is still somewhat fuzzy in the literature, but the area of study is still evolving. It can be safely said that the two are inter-related and that generally speaking scale free networks exhibit the clustering and short average path length of small world networks, but not all small world networks exhibit the power law distribution of scale free networks.

The implications of this new research into the structure of complex networks were very broad for a number of disciplines as varied as genetics, economics, molecular physics and sociology. One of the surprising findings was that not only did the World Wide Web fall into a scale free organization, but so did the Internet. The Faloutsos brothers (1999) found that the Internet followed power laws at both the router level and autonomous system (AS) level. The router level entails the fiber optic lines (edges) and the routers (vertices) that direct traffic on the Internet, and the AS level entail networks (AT&T, UUNet, C&W etc.- layer three transit providers) as vertices and their interconnection as edges. This meant that that the physical fabric of the Internet and the business interconnections of the networks that comprise the Internet both qualified as scale free networks. Before these discoveries, the Internet had been modeled as a distinct hierarchy or random network and the new finding had many implications throughout the field of computer science. Scale free theory and BA model have not been without debate.

Several arguments have been made stating that the BA model is too simplistic for the Internet and additional corollaries need to be made (Chen et al 2001). The re-wiring principle was one of Albert and Barabasi's (2000) responses to these criticisms, but overall the model has held. Tests of network generators based on power laws have been found to produce better models and efforts are being made to base new Internet protocols on these discoveries (Tangmunarunkit et al 2001, Radoslavov et al 2000). While these discoveries have paved the way for advancements in several fields, the question of the geography and location of these networks remain to be addressed.

**Network Evolution in the USA**

Analyzing spatial networks adds an additional variable into the problem, which increases the complexity of the issue. For one, networks can be planar or non-planar[4]. While this is not unique to spatial networks it does cause unique constraints. The majority of spatial networks that are based on Euclidean distance are planar. For instance, in a road network when two streets cross, generally speaking, you have an intersection. The number of edges that can be connected to a single node is limited by the physical space available to connect them. This fact makes the large number of connections needed for a power law distribution quite difficult to obtain. Even in non-planar spatial network such as airline networks the number of connections is limited by the space available at the airport, "such constraints may be the controlling factor for the emergence of scale-free networks" (Amaral et al 2000, p.11149).

---

[4] Planar network form vertices whenever two edges cross, where non-planar networks can have edges cross and not form vertices.

 It should be noted that Amaral et al (2000) did find that the airline network was a small world because of its small average path length, and other transportation networks such as the Boston subway has also been found to be small worlds (Latora and Marchiori 2001). It would seem that physical constraints prevent the formation of scale free networks in traditional transport networks, but is this true when one examines the transportation networks of an information economy. The information economy in part depends on fiber optic lines to transport digital goods and services. Fiber optic networks have a physical location and structure and can be analyzed as such.

This analysis will endeavor to do two things; (1) analyze the logical[5] US IP (Internet protocol) fiber optic infrastructure to determine if it forms a scale free network or not (2) explicitly encompass spatial aspects into a small world network model.    To examine the logical US IP fiber optic infrastructure data was collected for the years 1997-2000. 1997 and 1999 data was obtained from New York University's *Information Technology and the Future of Environment* project (SBR-9817778) (Moss and Townsend 2000), 1998 data was compiled from CAIDA's (Cooperative Association for Internet Data Analysis) MapNet application, and 2000 data was obtained from the University of Florida's *The Infrastructure of the Internet: Telecommunications Facilities and Uneven Access* project (BCS-9911222) (Malecki 2002). All four data sets cover the backbone layer-three transit providers[6] of the USA Internet and are very similar in composition. It should be noted that data in all three data sources is not always 100% accurate since carriers often advertise more bandwidth and lines than are actually in service and topological errors have been found in the past. These have been corrected for as best possible, and for the gross level of aggregate analysis in this paper these datasets are a viable information source.

---

[5] Logical network indicates that the connectivity matrix will be determined by how traffic is routed across the network using Internet protocol (IP).
[6] Refers to layer three of the OSI model described previously

For all four data sets total bandwidth connecting to a consolidated metropolitan area (CMSA) was tabulated. For the 1998 and 2000 data sets this was done through the construction of a matrix and the calculation of an accessibility index based on the bandwidth capacity of the links for each CMSA. For 1997 and 1999 data was provided with total bandwidth connected to the CMSA already tabulated. Capacity was totaled for each CMSA as the total number of mega bits per second (Mbps) of fiber optic connections to the CMSA, running IP. Since binary connectivity data was not available for 1997 and 1999 total capacity was utilized for comparison across the four years of data. Other researchers, including Amaral et al (2000) in his analysis of airline networks, utilize the weight of a link in their methodology to determine if a network is scale free. This approach is commonly used when structural network data is not available or the number of nodes is too small for a log-log plot. Utilizing capacity as measure of connectivity also makes sense since it takes into account the large number of lines connected to any one CMSA and the common practice of partitioning these lines. The vast majority of fiber optic partitions are as T-1's[7] that carry 1.544 Mbps, thus the Mbps total for each city can very roughly approximate the number of T-1 lines possibly available.

---

[7] A 1.45 Mbps (megabit per second) connection

The data for 1997-2000 was individually plotted as rank order distributions with log-log plots and fitted with a power law. For each graph the *y*-axis is the total bandwidth connected to a CMSA and the *x*-axis is the CMSA ranked in descending order. The results of each line fit can be seen in Figures 3-6 and the power law exponents and R squares are summarized in table 1. Prior research by Moss and Townsend (2000) found a high level of similarity in the exponential curves for the rank-size distribution plots by number of edges connected to a metropolitan area. The power law results in table 1 are calculated from $P[X = x] \sim x^{(k+1)} = x^{-a}$ where the exponent of the power law distribution is $a = 1+k$ (where **k** is the Pareto distribution shape parameter) (Adamic 2000). The exponent provides a rate of increase indicator, an exponent of -2 would indicate an increasing sequence of 1,4,9,16,25 or an exponent of 3 would indicate an increasing sequence of 1,8,27,64,125. As seen in table -1 the USA's backbone network has been incrementally increasing its power law exponent for each year, except for a small decrease from 1997 to 1998, but has also been increasingly moving away from a power law distribution. By 2000 the network's power law exponent is -1.82, approaching the range found in other real world scale free networks but the distribution is far from a power law. In fact the 2000 and 1999 data appears to be two different trends lines occurring. A closer examination of the data for 2000 reveals that there is a break between the top 110 CMSA's and the bottom 37 each with a distinctly different slope. Interestingly of the bottom 37, 33 do not have any of the high-speed 2.5, 5, or 10 Gbps connections.

The first distribution runs from the minimally connected locations at 45 Mbps and follows the power law until a tail forms, starting with Fort Pierce, FL with 4,976 Mbps and ending with Syracuse at 5,624 Mbps, consisting of 18 city vertices. The distribution then resumes in a normal power law trend to the top connected locations. The jump from Laredo, TX with 2,488 Mbps to Brownsville, TX at 4,976 appears at first glance to indicate a critical mass at which cities gain a level of preferential attachment into the network. Theoretically, as the USA Internet continues to evolve these kinks in the distribution will work out as connectivity spreads to more nodes, erasing clustered hierarchies. A closer examination of the data reveals that the reason for this clustering is a technology shock in the network. Beginning, for the most part, in 1999 several networks began provisioning dense wave dimension multiplexing (DWDM) lines with capacities of 2448 Mbps in their networks, a large increase in capacity from the more common 45 and 155 Mbps lines. A connection to two cities provides 4,976 Mbps and caused a whole cluster of cities to be bumped up into the 4,976-5,624 Mbps noted in the distribution. Massive investment in Internet backbone capacity has occurred between 1998 and 2000 in the US. In early 1998, only two of 38 national backbones offered bandwidth at OC-48 (2488 Mbps or 2.488 Gbps). By mid-2000, fully 17 of 41 backbone networks (41%) had installed capacity at bandwidths of 2488 Mbps or faster, as opposed to just 5% in 1998 (Gorman and Malecki 2002). Such bandwidths easily overwhelm networks of the slower capacity: a single OC-48 cable has the same bandwidth as 55 of the older DS-3 (45 Mbps) capacity. The current standard is OC-192, which moves data at speeds of nearly 10 gigabits per second, and work is underway to implement OC-768 (40 Gbps) in the near future.

The existence of a break and multiple slopes in the 2000 data could indicate that the diffusion of new high-speed technologies is not even across space and does not follow a power law. In order to test this assumption a binary connectivity distribution was built for 1998 and 2000 data, unfortunately this was not possible for the 1997 and 1999 data. The binary connectivity distributions can be found in figures 7 (1998) and 8 (2000). When bandwidth capacity is stripped from the network the trend reverses and from 1998 to 2000 to connectivity distribution is getting closer to a power law fit. The exponent is also increasing but the number is considerably lower than the bandwidth plots. This is most likely explainable by the way the binary connectivity distribution does not take into account all connections in the network, just if there is or is not a connection between cities. The reality of the network topology is most likely somewhere between the two, following a power law with an exponent higher than binary but lower than bandwidth. It would seem that when weighted links are used to examine scale free networks there is the possibility of shocks in the network, in this case a technology shock. Interestingly while the IP networks studied do generally follow a power law, the diffusion of new technologies across the network do not follow a power law geographically or topologically. This also leaves the question if applying curves, like power laws, is too simple an approach for weighted networks, but this falls out of the scope of this paper to answer.

The next condition for the BA model is preferential attachment. There is a higher probability for a new or existing node to connect or reconnect to a vertex that already has a large number of links than there is to (re)connect to a low degree vertex (Barabasi and Albert 1999). As the network grows incrementally it expands following preferential attachment. The probability (? ) that a new vertex will connect with another vertex (*i*) depends on the connectivity $k_i$ of that vertex so that $?(k_i) = k_i / S_j k_j$ (Barabasi and Albert 1999). Because of preferential attachment, a vertex that acquires more connections than another one will increase its connectivity at a higher rate; thus, an initial difference in the connectivity between two vertices will increase further as the network grows. This characteristic can be seen in the urban hierarchy of backbone connections. The Internet largely evolved out of Washington DC, through NSFNET and one of the original network access points, MAE East. Washington, DC has leveraged this historical preferential attachment to average the highest ranking over the four years of backbone connectivity data in the time series. While the rank order of the top ten cities has shifted (Table 2) they have consistently benefited from preferential connectivity to maintain the majority of connections in the network. Although it should be noted, that early first mover advantage for preferential attachment has succumbed to market size in many cases; the most obvious in the data being New York's move from sixth to first. Over the four time series the top ten cities have on average accounted for 57.4% of total bandwidth. The 1997-2000 time series appears to establish evidence of preferential attachment as one of many factors in the growth of the network. The actual testing of the BA equation to the time series was not possible since matrix connectivity data was not available for all four years. This is a future research avenue that could yield interesting results, especially in regards to predicting the future connectivity and growth of the network.

The last condition established by the BA model is re-wiring within the network. While this is not a feature of the network that can be tested explicitly it can be addressed anecdotally outside what has been cited in the literature. Re-wiring of the Internet occurs at many levels but at dramatically different rates. The backbone network, in general, operates at layer 3 of the OSI (open system interconnect) networking model. This is the layer where routing between networks occurs and rewiring within this virtual network occurs on a very frequent basis. Topologies and routes change frequently as new peering arrangements occur on one hand and the actually path of traffic changes constantly as congestion and traffic fluctuate on the other. The physical fiber that is installed in the ground is re-wired at much slower pace, but re-wiring does occur. Fiber into a city is typically leased from a carrier's carrier, like Enron, Williams, or Qwest. The long haul transit fiber into a city most often surfaces at a co-location facility, network access point, or a metropolitan area network interchange. At these junctures the individual conduits leased by multiple different backbone carriers are split off and run by various networks to their customer's locations. This allows for a considerable amount of fluidity in re-wiring topologies within backbone networks without actually digging up, turning off, or laying new fiber. The most dramatic example of this type of re-wiring was the change in Cable and Wireless's network when they acquired MCI's network. The network was significantly re-wired from a star topology focusing on connectivity to coastal cities to a partial mesh topology concentrating connectivity to interior vertices (Gorman and Malecki 2000). While the re-wiring principle occurs at various levels of the data examined and at different rates it is very much a factor affecting the distribution and connectivity of the network.

**A Spatial Small World Model**

The results of the first step of this analysis establish that the US Internet infrastructure is evolving towards a scale free network according to the methods devised by statistical physics.  The problem with this model of analysis is that it only looks at the US network as a rank order distribution.  Rank order distributions that follow power laws have been found in geographic phenomenon before.  The most significant finding being city size, which follow a Zipf (1949) distribution, but firm size has also been found to follow similar distributions as well as host of natural phenomenon such as earthquake size (Amaral et al. 2000).  In fact several Internet related variables follow similar power laws, especially at the global level by country.  Utilizing data from Telegeography (2002) Internet users, domains, and cross border bandwidth were analyzed (Table 3). The rank order distribution of country connectivity produces a power law exponent of 2.5, for number of Internet users the power law exponent is 2.6, and for number of domains located in a country 2.8.  Domains and users, though, are not networks, they are simply agglomerations and each node (user or domain) is independent of all other nodes.  This is not true when looking at fiber networks where nodes are interdependently linked to each other.  A common explanatory variable for Internet use has been wealth, often accounted for by differences between GDP in countries.  A least squares analysis verifies this assumption – when Internet users per country is regressed as the dependent variable and GDP per capita as the independent variable, GDP explains 71% of the variation in Internet users between countries.  When this same analysis is run and international bandwidth is substituted as the dependent variable only 20% of variation is explained by GDP per capita. There is more going on in a geographic network than a simple rank-order distribution of wealth and size. To develop a spatial small world model the geography of the network needs to be taken into consideration.

The fundamental idea behind small world networks is the notion that networks gain efficiency by having a large number of local links and a few global links connecting local clusters together. Local and global are geographic concepts, and the basic mathematical concept of small worlds can be adjusted to fit a spatial framework. Watts (1999) explicitly examines spatial graphs and the role of Euclidean distance in his analysis of small world networks and found that, "It appears, in fact, that spatial graphs with more exotic distributions do display small world features, but the matter is formally unresolved" (Watts 1999, p.42). This analysis will not try to resolve the problems of Euclidean distance, but instead will examine the problem from a different perspective utilizing geographic regions instead of distance.

To do so a binary matrix of the US Internet infrastructure was utilized as a test case. The United States was divided into the four census regions (South, West, Midwest, Northeast) and each city in the Internet infrastructure matrix was assigned to a census region. Census regions were an appropriate choice of geographical units for this analysis because of their direct connection to population distribution. Significant analysis has been done on several of the Internet infrastructure variables in regards to population. Malecki (2001) found that the most significant explanatory variable of bandwidth is population, and similar results were found with colocation facilities (Malecki and McIntee 2002). These findings were further reinforced at the global level when the Internet router network was examined, finding that the physical layout of nodes form a fractal set, determined by population density patterns around the globe (Yook et al 2001). A similar study at Boston University found the same effect when population was controlled for with economically homogeneous regions (Lakhina et al 2002).

Utilizing census regions the following procedure was established, for each city the number of local links to other cities in the same census region were totaled along with the number of global links connecting to cities in other census regions. From this data the following approach was developed to identify cities that act as the super connected nodes that provide the key global connection in the network:

Consider a large network of $n$ nodes, spanning an area $A$ consisting of $m$ regions, with variable number of nodes inside each region that have variable number of connections from each region to other regions. For a region $r$ with $p$ number of nodes, a $p \, x \, p$ contiguity matrix represents connections between these nodes. Then, one could construct a contiguity or adjacency matrix for the entire network of $m$ regions, as a block diagonal matrix, where matrices along the main diagonal refer to the contiguity matrices for each of the regions, while interregional connections are represented as the off-block-diagonal elements. Let $M$ denote such a matrix (Table 9).

If a node *i* in region *r* is connected to another node *j* in the same region, then that connection is considered as a local link and is denoted by $q_{i(r)j(r)}$, where the value is one if a link exists and zero if it does not. On the other hand if node *i* in region *r* is connected to node *l* in region *s* then that connection is considered as a global connection and is denoted by $g_{i(r)l(s)}$, where the value is one if a link exists and zero if it does not. Thus, in theory, one may associate with each node node *i(r)*, a global connectivity index as a ratio between its global and local connections, weighted by the total number of global and local connections for the entire network.

The total number of global connections *G* is computed from the elements of the block upper triungular matrix of M, of *m* regions, each with variable number of nodes:

$$G = \sum_{i(1)} \sum_{s>1}^{m} \sum_{l(s)} g_{i(1)l(s)} + \sum_{i(2)} \sum_{s>2}^{m} \sum_{l(s)} g_{i(2)l(s)} + \ldots + \sum_{i(m-1)} \sum_{s>m-1}^{m} \sum_{l(s)} g_{i(m-1)l(s)} \qquad (1)$$

Note that, since *m* is the last region in the block diagonal matrix, its global connections have already been computed in the previous *m*-1 blocks.

The total number of local connections *L* is a sum over all the local connections over *m* regions and is given by:

$$L = \sum_{i(1)} \sum_{j(1)>i(1)} q_{i(1)j(1)} + \sum_{i(2)} \sum_{j(2)>i(2)} q_{i(2)j(2)} + \ldots + \sum_{i(m)} \sum_{j(m)>i(m)} q_{i(m)j(m)} \qquad (2)$$

Then the global connectivity index for a node *i* in region *r* is given by:

$$C_{i(r)} = \left( \frac{\sum\limits_{s \neq r}^{m} \sum\limits_{l(s)} g_{i(r)l(s)}}{1 + \sum\limits_{j(r), j \neq i} q_{i(r)j(r)}} \right) \times (G + L) \qquad (3)$$

The numeral of 1 in the denominator indicates a self-loop of a node.

The ratio of global links to local links provides an indicator of how well the city acts as a global connector in the network and the weighting of the scores by the total number of links balances the measure with the overall connectivity of the node. To test the model the equations were run with the 2000 Internet infrastructure data and produced the results seen in Table 4. For each city equation 1 produced an indicator of the number of global connections, equation 2 produced an indicator of the number of local connections, and equation 3 provides a ratio of global to local connection weighted by the total connectivity of the city. Table 4 provides the results of equation 3 and then ranks the scores from highest to lowest for the top twenty-five cities. Running the model produced an interesting distribution of city rankings, each region (South, West, Midwest, Northeast) placed a city in the top four and then another city for each region in the top eight. Each region produced a first tier and second tier global connector at highly similar levels of connectivity. These cities serve as global hubs that connect the cities in their regions to the rest of the network. The links in the diagonal of the regional matrix totaled 751 and the off diagonal global links totaled 327, illustrating the small world character of the network – a large number of local connections with a smaller number of global links connecting clusters.

This is a markedly different result that when just the level of connectivity of a city is accounted for, seen in Table 5. The top city by level of connectivity is Atlanta and New York is only ranked sixth. Only two of the top four remain in the top four, Chicago and San Francisco. From a purely geographic view New York is not well located to act as network hub. It is positioned at the edge of the network and is not conveniently positioned to connect to multiple proximal locations. This is where the networks of the information economy diverge significantly from traditional industrial networks like rail and roads. Information networks, like the Internet, are non-planar. This allows for the global links of small world network to be formed. New York can connect directly to San Francisco completely bypassing all intervening cities between the two.

One of the key aspects of this approach to examining networks is the selection of regions. In the results described above a few large regions were chosen. In order to test the effect of region size on the model the same data was run but utilized Census divisions instead of Census regions. Census divisions break up the United States into New England, Middle Atlantic, South Atlantic, East North Central, East South Central, West North Central, West South Central, Mountain and Pacific. This increased the number of regions analyzed from four to nine. The results of running the model for the top twenty-five CMSA's is presented in table 6. The most notable difference is that St. Louis jumps from number eight in the ranking to number three. This jump most likely occurs because St.

Louis is located in the West North Central region, which is sparsely populated, containing North Dakota, South Dakota, Nebraska, Iowa, Kansas, Missouri, and Minnesota, and there are not many local places in the region to connect to. In fact, St. Louis only has three connections within its region, but 17 connections outside its region. St. Louis plays an important role as a relay hub between the East and West coast, but local connection within the region are handled by Kansas City and Minneapolis that collectively have eleven local connection within the region[8]. In general though the choice of smaller regions did not have an adverse effect on the model. When a Pearson R correlation was run for the results of both models there was a 93.55% correlation. Further, the same trend of one global connector in each region was found with the top eight cities each coming from a different region. The only exception was the East South Central Region (Kentucky, Tennessee, Alabama, and Mississippi), which did not have a city show up until Louisville, KY at number twenty in the ranking. The most likely reason for the underperformance of the region is the lack of a CMSA with a large enough market to attract global connections from other large market nodes. The results of the analysis point toward Atlanta serving as the proxy for the region's global connector, this would also complete the one global connector from the top nine in each region, since Atlanta is ranked ninth.

One of the drawbacks of the model is that cities that are on the border of a region can have artificially inflated numbers from connection to local neighbors in adjacent regions. To examine the effect of the border phenomena all cities that were located within 200

---

[8] St. Louis's two local connections are to Kansas City and Minneapolis respectively, reinforcing the hubbing effect in the region.

miles of a border were assigned a dummy variable in a regression model that included

CMSA population, bandwidth, and wireless auction value. The dummy variable

accounted for 1.5% of variation in the model, which seems small enough for the model to

be a viable tool for analysis. This shortcoming could be removed by basing the model on

Euclidian distance between cities, but this approach goes outside the scope of this paper.

A third approach to modeling spatial small worlds could be the use of a local and global Moran *I*

statistic to identify global and local nodes in the network.

**Nodes Defining Regions Versus Regions Defining Nodes**

The method outlined above identifies what nodes are serving as hubs for a predefined region, but the definition of region can be somewhat arbitrary. The choice of regions for the analysis will invariably have an effect on the results, and this can be a desired effect depending on what is justified by the theoretical assumptions and existing literature.

Further, if a study endeavored to identify key hubs in an existing economic or political region using those boundary definitions would be a necessity. Another possibility is to define regions based on a variable that has been highly correlated with the network being analyzed. In the case of this study the high correlation in several studies in the literature of information infrastructure and population made the choice of census regions (based on population) appropriate.

There is, though, another approach the problem, instead of using regions to define nodes,

use nodes to define regions. A substantive literature has been developed around the concept of developing regions or sub-groups from a network, especially dealing with social networks. While this literature will not be covered in depth three techniques appropriate to the problem, LS sets, Lambda sets, and hierarchical clustering, will be covered regarding their fit for a possible solution.

Social network analysis structures the relationships between people as graphs, where people are vertices and the relationships between them are edges. Sociologists will often analyze these networks to identify sub-groups or cliques within the network. Two methods of identifying sub-groups in the network are LS and Lambda sets. The LS set defines subgroups as a set of vertices that have a greater number of connections between subgroup members than connections to members outside the subgroup Wasserman and Faust). Siedman (1983) defines this mathematically as, "A set of nodes $S$ in a social network is an LS set if each of its proper subsets has more ties to its complement within $S$ than to the outside of $S$ (p. 98)." While this approach is suitable for social network analysis it runs into problems in a small world or scale free context. Both concepts, small world and scale free, are based on hub nodes connecting groups of poorly connected nodes to other hub nodes. Thus, it is very unlikely that the nodes connected to a hub would have many connections to each other. A similar problem is encountered in the Lambda set. A Lambda set is a similar idea to an LS set, but is based on the idea that a sub-group should be hard to disconnect. A set of nodes is a Lambda set "if any pair of nodes in a Lambda set has larger line connectivity than any pair of nodes consisting of one node from within the Lambda set and a second node from outside the Lambda set (Wasserman and Faust p.270)." The concept of defining sub-groups as a cluster that is hard to disconnect runs counter to empirical work done on scale free networks. Albert et al (2000) found that disconnecting hubs from a scale free network cause the poorly connected vertices to be disconnected resulting in a rapid degradation of the network. Applying either the LS or Lambda set to a scale free network would result in simply defining a the set of hubs that are highly interconnected with each other, and is unlikely to uncover any spatial clustering that could be identified as regions.

An alternative approach is to use hierarchical clustering, where vertices are placed into subsets that are structurally similar. Structural equivalence is dictated by a distance measure $d_{ij}$ between vertices $i$ and $j$ that is based on a threshold a. Vertices $i$ and $j$ are considered structurally equivalent when $d_{ij} > a$. The threshold a is often determined by the clustering coefficient for the network $c_{ij} = a$ (Wasserman and Faust 1994). Utilizing the methodology outlined above a clustering coefficient was calculated for the 2000 network data set using UCINET (Borgatti et al 2002). The resulting clustering coefficient was .508, thus $d_{ij} > .508$. UCINET was used again to partition the network into clusters with $d_{ij} > .508$ and resulted in ten distinct clusters mapped in table 10. Clusters one and three are quite small each only having two cities but both located geographically proximate to each other, and several distinct geographic regions emerge among the remaining eight clusters. Cluster six forms a very distinct region in the northeast and cluster ten illustrates a strong west coast bias anchored by the east to west coast hubs of Chicago and Kansas City. Cluster seven sits largely in the mid-west with a strong connection spanning to the Pacific Northwest. Just south of the northwest seven cluster is the smaller cluster two containing four cities along the northwest coast. Cluster nine is small but encompasses most of west Texas. Cluster eight is the largest accounting for the rest of Texas and most of the inland southeast. The southeast is the most confusing region, interspersing cities from region eight, four and five. Region four accounts for most of the southeastern coastal areas including all of Florida and region five spans southern Appalachia reaching into the Ohio Valley. There are also a few odd outliers like Charlottesville, VA that is part of the strongly west coast cluster ten. Hierarchical clustering is based solely on network structure, so a few outliers would be expected, but the formation of several distinct geographic clusters is evidence of a strong spatial factor in the formation of networks.

While hierarchical clustering does provide evidence of a strong spatial component it does not directly address the problem if the hubs in a scale free networks form distinct geographic regions. To directly address this issue an algorithm is proposed that assigns each node in a network to a hub and then examines the results for evidence of clustering.

Algorithm to generate regional domains

1. For a network N of $n$ nodes, generate an adjacency matrix **A** and distance matrix **W**. The members of matrix **W** represent the physical distance between any two nodes of network N.

2. Next compute the shortest paths for each node of N using adjacency matrix **A**

3. Let the maximum number of hops between the shortest-longest path be $H$. By definition, the minimum number of hops in a shortest path between any two nodes is 1.

4.   Assuming that the adjacency matrix **A** is symmetric, compute either egress or ingress connections $c(i)$ for each node $i$ of N.

5.   Rank in descending order these nodes by ingress (egress) connections $c$.

6.   Create a set $m < n$ of an arbitrary number of top ranked nodes computed in step 5.

7.   For each member $j$ of set $m$ compute lists $L_r(j)$, of nodes that are 1,2, … $H$ hop distant from node $j$ and  $r \varepsilon$ [1,2,..$H$].

8.   $R_j = \sum_r L_r(j)$, represents a region around node $j \varepsilon m$.

9.   Starting with the highest ranked node of set $m$, compare the list $L_r(j)$ to $L_r(j?)$, of each of the remaining nodes of set $m$.

10. Do not include one-hop connections (if there are any) between the top $m$ nodes.

11. If there is a common node $q$ that is $h$ hops away from both $j$ and $j?$, then, compare the physical distances $d_{jq}$ and $d_{j?q}$ between nodes $j$ to $q$ and $j?$ to $q$ from the distance matrix **W**.

12. If $d_{jq} <= d_{j?q}$ then node $q$ belongs to the list $L_r(j)$ or region $R_j$,whose members are exactly $h$ hops away from node $j \varepsilon m$.

13. If $d_{jq} > d_{j?q}$, then $q$ belongs to the list $L_r(j?)$ region $R_{j?}$,whose members are exactly $h$ hops away from node $j? \varepsilon m$.

When the algorithm was run with the 2000 dataset the regions depicted in table 11 were produced.  Cities that are one hop from the regional hub are given the hubs abbreviated name (i.e. ATL = Atlanta) and cities that are more than one hop away are designated by the abbreviated name followed by the number of hops  (i.e. ATL2 = two hops away from Atlanta).  The regions produced by the algorithm are more distinct than the hierarchical clustering technique. This is most likely due to the shortest hop approach and the inclusion of a distance variable to break ties between places with the same shortest paths to two or more hubs. It should be noted that the distance variable could be substituted with a bandwidth capacity variable or other variable of the researchers choice as best fits the algorithm's application. In this case distance was used because network design most often incorporates a distance cost variable when selecting link build outs (Cahn 1999).

Overall the algorithm's output illustrates several strong regions. The most prominent being the Chicago region, again illustrating the strong connection Chicago has as a hub to the West. Atlanta also forms a very prominent region, incorporating the largest number nodes into its region. San Francisco's region dominates the West coast leaving Los Angeles with only the southern tip of California and areas of the lower Southwest. Denver and Dallas subsume the rest of the West with distinct regions of their own. The sole outlier of the map is Boise's inclusion in the Denver region, transcending the boundary of the Chicago region. The Kansas City region includes the parts of the Midwest not falling under Chicago, and lastly the Northeast and Mid Atlantic are split by New York and Washington DC respectively.

From a small world perspective the algorithm illustrates the strong spatial bias that global connectors have towards the nodes that are local to them in the network. The nodes in each region are not only local to their hubs in terms of network hops but also in terms of spatial proximity. Further, the algorithm illustrates specifically that the poorly connected nodes of scale free networks can form spatially distinct clusters to their associated hubs.

The results reinforce the spatial nature of complex networks found in the "region defining nodes" approach.

## The Resurrection of Distance?

The friction of distance in an information network is greatly reduced in comparison with that of traditional industrial networks but not eliminated. Global connectors will connect to each other largely regardless of distance, but smaller cities will connect to the closest global connector. Building out the fiber optic links between cities is not cheap, and prospective over-building has been largely attributed as the cause of the current economic slump in the telecommunications sector (Economist 2001). While laying fiber is cost prohibitive, relative to the cost of traditional infrastructure networks it is far less inexpensive. This allows, when sufficient economic demand dictates, for direct long distance connections to be made between markets of sufficiently large size.

To do a brief test of this assertion the model was run against bandwidth, population, and auction value for each metropolitan area. Bandwidth is an account of the total amount of fiber capacity connecting the city to other regions for 2000, population is from the 1999 census, and auction values is an aggregated number from the Federal Communication Commissions' auctions for personal communication service (PCS) spectrum blocks. PCS auction values for metropolitan areas are highly correlated with the market size, especially in relation to information services (Gorman and McIntee forthcoming). The results in table three show a statistically significant relationship between all three variables and the small world model. The results reinforce the assertion that the larger the market for information services the more likely a global connection will be established between two cities. Bandwidth produced the strongest relationship with an $R^2$ of .774 and the relationships between global connections and market size is in line with existing research on Internet infrastructure and global cities (Malecki 2002, Townsend 2001, Beaverstock et al 2000). This finding is further confirmed by recent statistical analysis of bandwidth agglomerations in US metropolitan areas that found globalization, knowledge jobs, economic dynamism, population, innovation capacity, and high tech clustering act as a significant explanatory model (Malecki 2001).

**Security Implications of a Spatial Small World**

The spatial small world and scale free structure of the US Internet has implications beyond new geographic patterns of connectivity. Scale free networks are very fault tolerant but very susceptible to attack (Albert et al 2000). Research has found that a scale free network model remains connected when up to 80% of nodes are randomly removed from the network. On the other hand, when the most connected nodes are removed the diameter of the network increases rapidly, doubling its original value if the top 5% of nodes are removed (Albert et al 2000). In fact, when the router network was examined when only the top 2.5% of nodes where removed the network's diameter tripled (Albert et al 2000). The emergence of scale free characteristic in the spatial networks studied in this analysis illustrate the need to look more closely at the security implications of current spatial network structure. A benefit of this analysis over previous research is the ability to provide a physical location for critical nodes in the network. Another area where work could be improved is incorporating network interconnection into these models. Previous work has looked at the interconnection patterns at the AS (autonomous system) level of connectivity where location is meaningless and the router level where location is meaningful, but the two have not been looked at concurrently. The ability to model the Internet at a more accurate operational level should produce results more applicable to the reality of the network. These will form future directions in our research.

**Competition and Network Efficiency**

The market driven small world structure of the US Internet has not only created a very fault tolerant network, but also a highly efficient end-to-end communication services seen in the low diameter of the network. Historically, "the avoidance of monopoly rents and the need to ensure continuity and quality of supply have been among the key drivers for the creation of new trade routes" (Paltridge 2002, p.2). While this study was restricted to the United States it creates a methodology very well suited to performing a global analysis. The definitions of global and local links in the model are fluid to changes in scale. For the case of a global analysis instead of census regions continents or countries could be used to define what is global and what is local. Since the Internet at the router level has been determined to be scale free, changing the geographic scale of analysis should not theoretically change the network structure (Faloutsos 1999). While theoretically this would be expected the realities of different levels of regulation and competition across different countries and jurisdictions makes this assertion tenuous. Comparisons of the USA's, Europe's, China' network infrastructures found wide disparities in network structure with only the USA's network forming a scale free structure (Gorman 2001). How these disparities will affect the aggregated global network remains to be seen, but is a future direction of this research. Further, the simple spatial small world model presented is also not only restricted to the analysis of Internet infrastructure. Any network that can be decomposed into links, nodes and geographic location can be analyzed. Studies have found many similarities in the network structures of Internet infrastructure and airline routes (Choi et al 2001).

## Conclusions

The findings of the research presented in this paper could provide the beginning of a network theory for cities. Batty (2001) calls for a small-worlds theory of cities, and the findings of this simple spatial small world model are a small step in that direction. The initial findings of this analysis point to a strong spatial component to existing complex network theory. Further, the findings point to new patterns of how cities are being wired in the information economy. Although the urban hierarchy has not significantly changed the way these cities transport goods and services, the form of information exchange has. Information networks allow direct global connections between distant places at speeds and volumes not possible before. While this interconnects the global economy in a tighter weave, geography and distance are still important factors. The location of super connected nodes has important economic, business and security implications. Much work remains to be done in expanding the scope of study, sophistication of techniques, and examining causal socio-economic factors. The model also comes up short in dealing with the vagaries of where boundaries are placed in the model. Locations that are close to one or multiple boundaries have an explicit advantage in the ranking algorithm. This makes for good results at the bottom and top of the ranking but muddies the picture for locations ranked in the middle. There is still much to be gained from the advanced work done by statistical physicists in the field. The mechanisms that generate faster growth of the most connected vertices in the BA model are identical to the city growth model of Simon and Bonini (1958) (Amaral et al 2000). Further, the addition of a spatial perspective opens new questions for the existing literature on complex networks. Finally to truly test the model the area of study needs to be global. This is an avenue that is currently being explored through data collection for a worldwide fiber optic database.

The time-series data presented in this analysis presents an opportunity to look at how networks evolve over time. While the number of nodes involved is not significant enough to test many of the theoretical aspects of scale free phenomena some interesting results present areas of further investigation. The BA model predicts that independent of time and continuous growth networks will organize themselves into a scale-free stationary state. Meaning that the exponent of the power law distribution should not change significantly as new vertices is added to the network. The finding of the time-series data does not appear to support this prediction as the exponents change significantly each year. One possible explanation of this deviation is the role of technology shocks to the network. The BA model is built based on binary contiguity connectivity matrices, where as the matrices in this study used were bandwidth-weighted matrices. When new telecommunications technologies, like optical networking, are introduced into the network connectivity can radically change causing spikes in the connectivity distribution (see 1999 and 2000 power law distributions of the USA Internet). The explanatory level of technology shocks and their effect on the future connectivity and distribution of the network needs to be further analyzed.

The methodologies and finding from this analysis also pose some interesting new approaches to current policy issues. Joining the growing global economy through networked connectivity is a policy prescription that is gaining increasing promotion from many sources, ranging from local economic development agencies to the World Bank. The evidence to support the benefits of wiring into a global network is persuasive. The sum of OECD countries that have deregulated their telecommunications market have enjoyed increased international connectivity, increased competition, decreased prices, a broader variety of services, and economic growth in the sector (Paltridge 2002). The roles of these policies and their interplay with the issue of the digital divide remains to be answered. Applying small world and scale free theory to these and other geographic issues appears to be a promising perspective to investigate evolving aspects of the information economy.

**Table 1. Power Law Distribution Results for the USA Internet from 1997-2000**

| | |
|---|---|
| 1997: | $y = 62702x^{-1.5953}$ |
| | $R^2 = 0.9213$ |
| 1998: | $y = 104628x^{-1.5443}$ |
| | $R^2 = 0.9015$ |
| 1999: | $y = 1.7558x^{1.8087}$ |
| | $R^2 = 0.8398$ |
| 2000: | $y = 9E+06x^{1.8252}$ |
| | $R^2 = 0.6879$ |

**Table 2. Top 10 CMSA Change in Rank 1997-2001**

| CMSA | 1997 (Mbps) | 1998 (Mbps) | 1999 (Mbps) | 2000 (Mbps) |
|---|---|---|---|---|
| New York | 6766 (4) | 9543 (5) | 22232 (6) | 234258 (1) |
| Chicago | 7663 (2) | 14809 (2) | 23340 (4) | 221738 (2) |
| Washington | 7826 (1) | 14174 (3) | 28370 (1) | 208159 (3) |
| San Francisco | 7506 (3) | 14924 (1) | 25297 (3) | 201772 (4) |
| Dallas | 5646 (5) | 10985 (4) | 25343 (2) | 183571 (5) |
| Atlanta | 5196 (6) | 5426 (8) | 23861 (5) | 149200 (6) |
| Los Angeles | 5056 (7) | 9397 (6) | 14868 (7) | 140649 (7) |
| Seattle | 1972 (9) | 5409 (9) | 7288 (10) | 109510 (8) |
| Denver | 2901(8) | 5942 (7) | 8674 (9) | 97545 (9) |
| Kansas City | 1080 (10) | 2715 (10) | 13525 (8) | 89292 (10) |

**Table 3. Analysis of Global Internet Data (2001)**

| Country Level Variable | Power Law Exponent | $R^2$ |
|---|---|---|
| Domains | -2.7575 | .8229 |
| International Bandwidth | -2.4595 | .6284 |
| Users Online | -2.4786 | .77 |

**Table 4. Global connectivity: top-25 consolidated metropolitan statistical areas (CMSAs) by small-world rank, utilizing census-region data (2000)**

| Region | CMSA | Small World Rank |
|--------|------|------------------|
| NE | New York | 59.1 |
| MW | Chicago | 52.4 |
| W | San Francisco | 48.6 |
| S | Washington | 43.1 |
| NE | Boston | 40.8 |
| S | Dallas | 40.2 |
| W | Denver | 35.2 |
| MW | St. Louis | 24.4 |
| MW | Cleveland | 19.1 |
| S | Louisville | 18 |
| MW | Kansas City | 17.1 |
| W | Phoenix | 15.1 |
| W | Seattle | 15.1 |
| W | Los Angeles | 14.8 |
| S | Atlanta | 14.7 |
| W | Salt Lake City | 13.8 |
| MW | Tulsa | 13.3 |
| MW | Indianapolis | 12.4 |
| S | El Paso | 11.3 |
| MW | Detroit | 10.5 |
| S | Houston | 9.9 |
| NE | Philadelphia | 9.3 |
| MW | Cincinnati | 8.6 |
| S | Austin | 8 |
| W | San Diego | 6.3 |

**Table 5. Top Twenty-five CMSA's by Connectivity (2000)**

| CMSA | Binary Links |
|------|--------------|
| Atlanta | 44 |
| Chicago | 41 |
| San Francisco | 37 |
| Dallas | 36 |
| Washington | 34 |
| New York | 28 |
| Denver | 26 |
| Houston | 24(t) |
| Kansas City | 24(t) |
| Los Angeles | 23 |
| Cleveland | 21 |
| St. Louis | 20 |
| Salt Lake City | 19 |
| Boston | 17(t) |
| Phoenix | 17(t) |
| Seattle | 17(t) |
| Indianapolis | 16 |
| Miami | 15 |
| Detroit | 14(t) |
| Orlando | 14(t) |
| Charlotte | 13 |
| Cincinnati | 12(t) |
| Columbus, OH | 12(t) |
| Jacksonville | 12(t) |
| Minneapolis | 12(t) |
| New Orleans | 12(t) |

t = tie in rank

**Table 6. Top Twenty-five CMSA's by Small World Rank Using Census Divisions (2000)**

| REGION | CMSA | Small World Rank |
|---|---|---|
| Mid Atlantic | New York | 134.4 |
| Pacific | San Francisco | 103.6 |
| West North Central | St. Louis | 85 |
| East North Central | Chicago | 82 |
| West South Central | Dallas | 75 |
| Mountain | Denver | 74.3 |
| New England | Boston | 59.5 |
| South Atlantic | Washington | 51 |
| South Atlantic | Atlanta | 50.3 |
| Mountain | Salt Lake City | 44.3 |
| West North Central | Kansas City | 36 |
| Pacific | Seattle | 34 |
| Pacific | San Diego | 33 |
| Pacific | Los Angeles | 32.2 |
| West North Central | Minneapolis | 27 |
| Mountain | Phoenix | 26.7 |
| West South Central | Houston | 26 |
| East North Central | Cleveland | 25.2 |
| East North Central | Indianapolis | 22.9 |
| East South Central | Louisville | 21 |
| Mountain | Boise | 15 |
| East South Central | Jackson MS | 15 |
| West South Central | New Orleans | 14 |
| East South Central | Memphis | 13.5 |
| West North Central | Joplin (t) | 11.7 |
| South Atlantic | Miami (t) | 11.7 |

**Figure 1.**

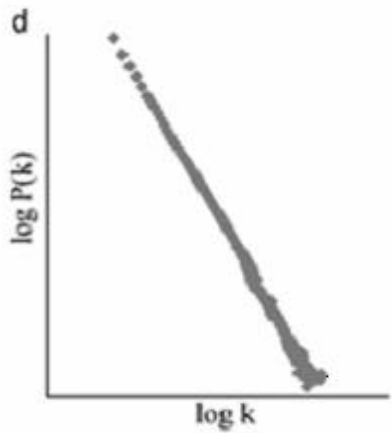

A plot of the logarithm of the probability that a vertex in the network interacts with *k* other vertices [P(*k*)] against the logarithm of *k* (source: Barabási, 2001).

**Figure 2**

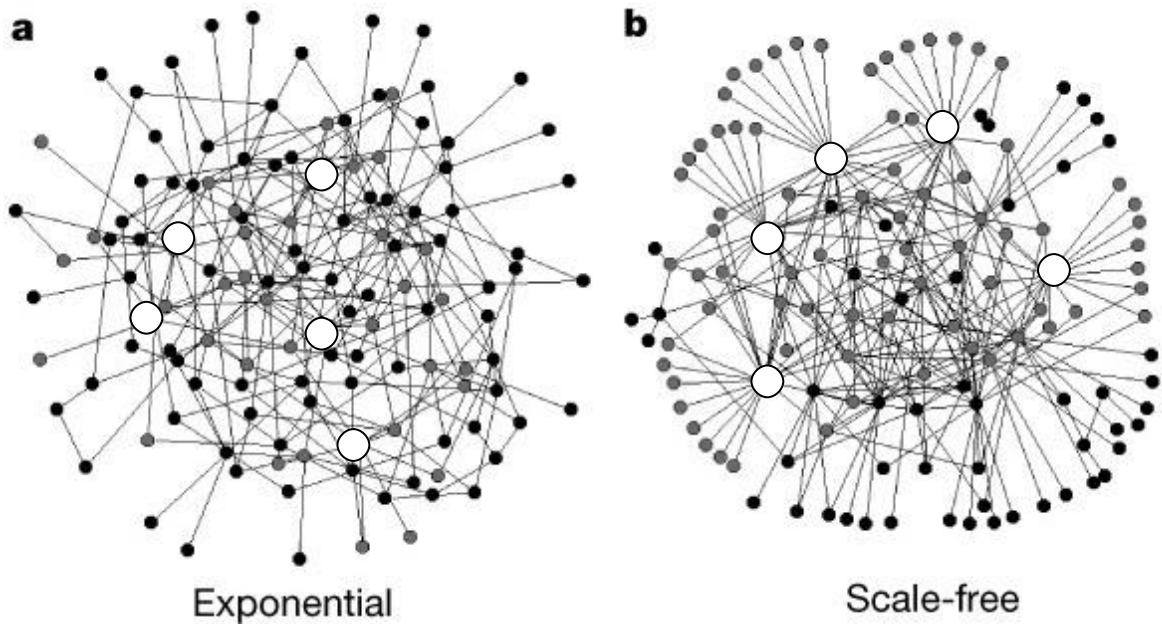

Figure 2. Comparison of (a) the exponential network model of Erdös and Renyi (1960) with (b) a scale-free model.  Note: Red dots are the five nodes with the highest number of links with green dots being their first neighbors. In the exponential network only 27% of the nodes are reached by the five most connected nodes, in the scale-free network more than 60% are reached, demonstrating the importance of the connected nodes in the scale-free network Both networks contain 130 nodes and 430 links (*k* = 3.3).

Source: Barabasi (2001)

Figure 3 & 4.

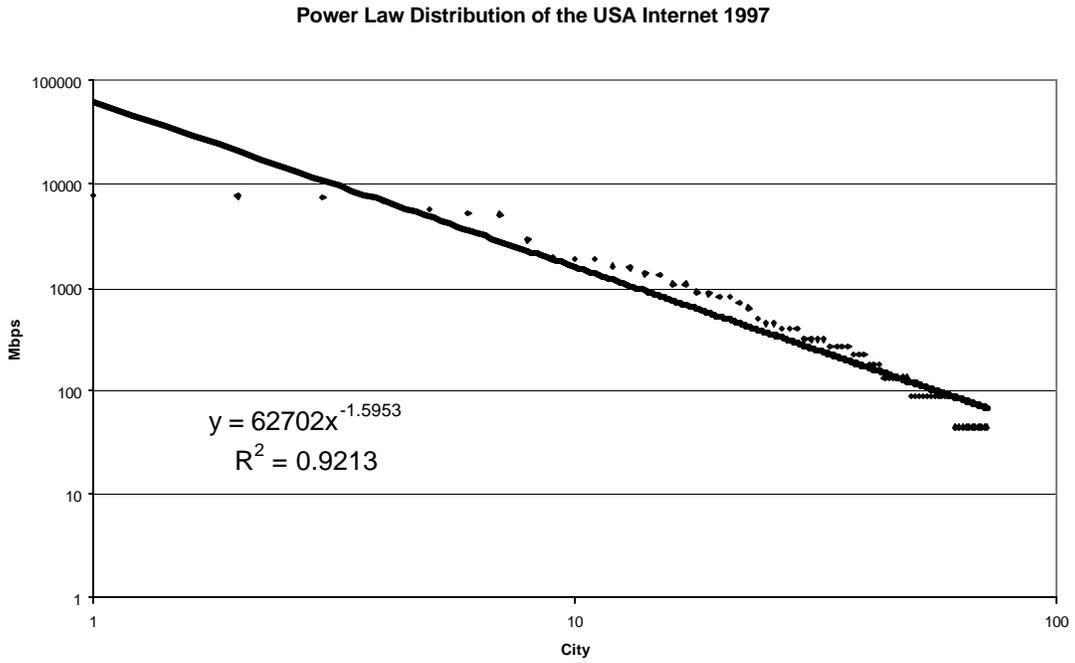

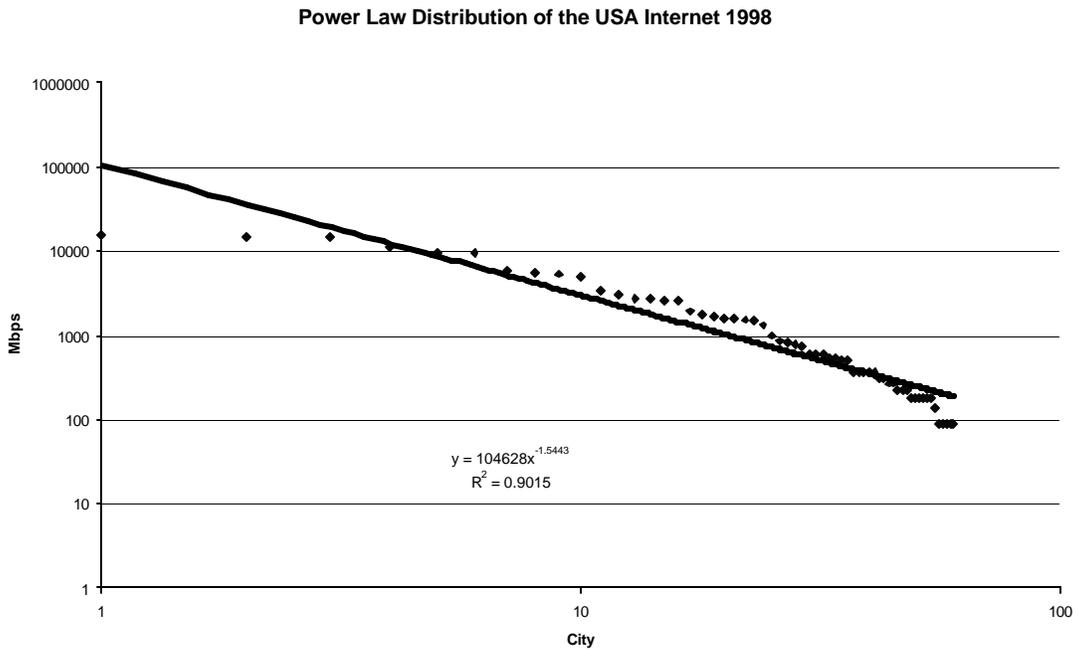

Figure 5 & 6

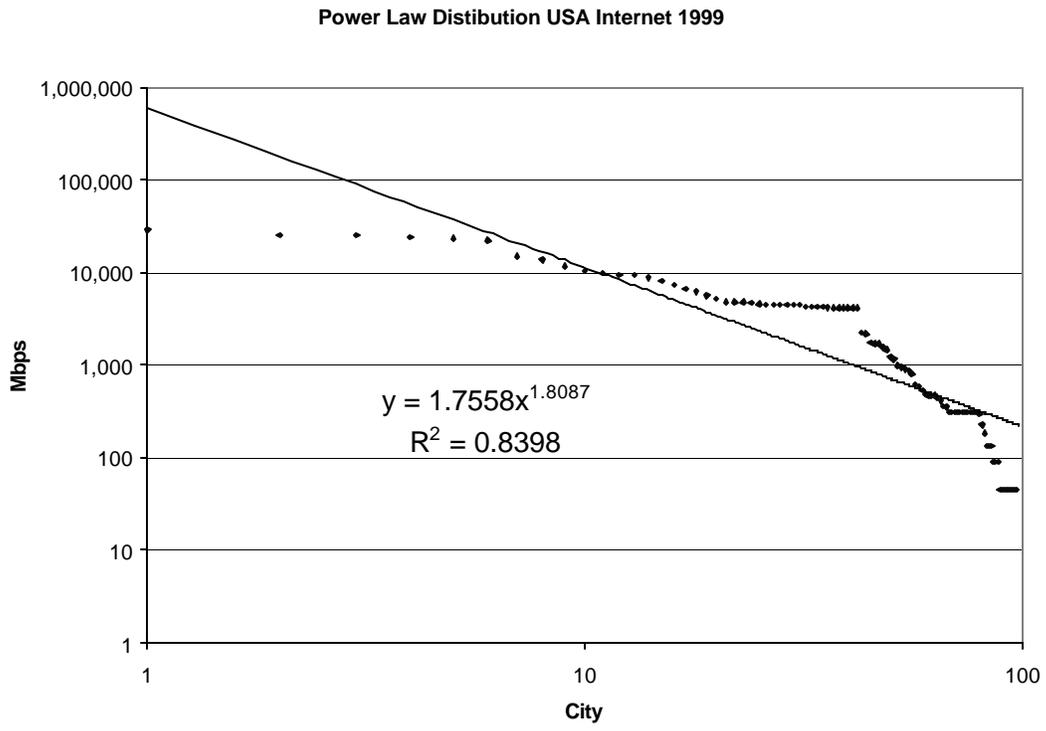

**Power Law Distibution USA Internet 1999**

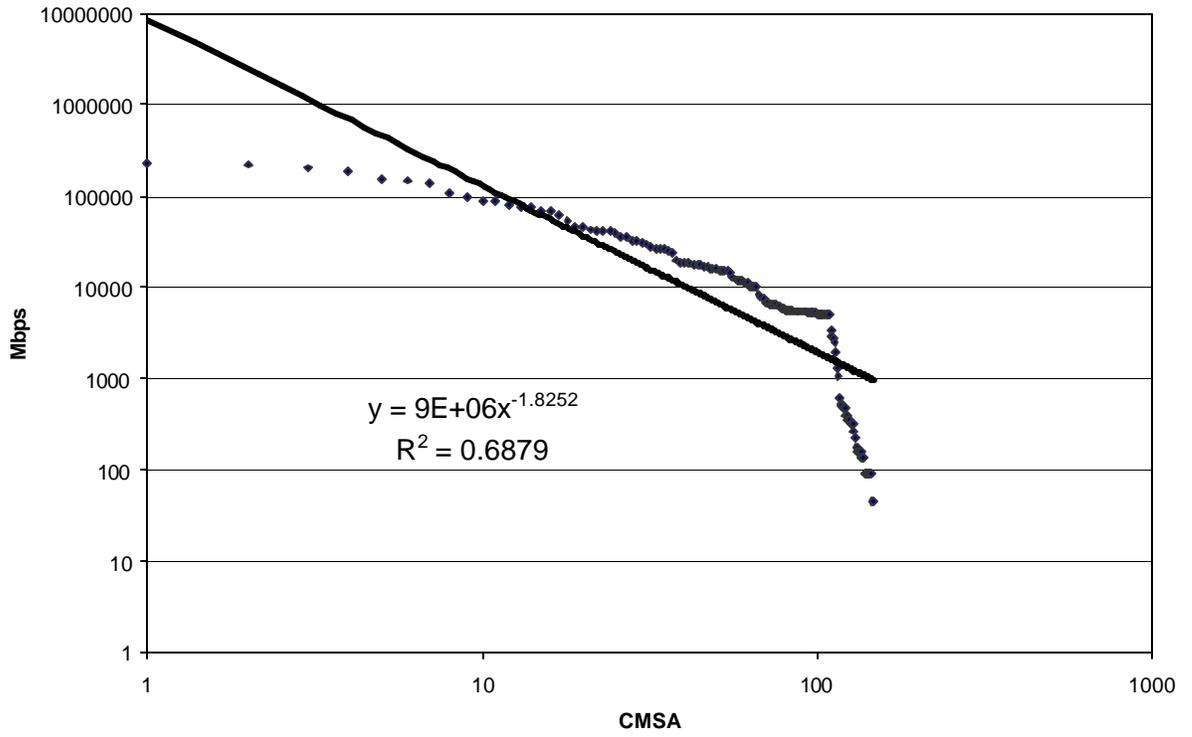

**Power Law Distibution of the USA Internet 2000**

$y = 9E+06x^{-1.8252}$

$R^2 = 0.6879$

Figure 7 & 8.

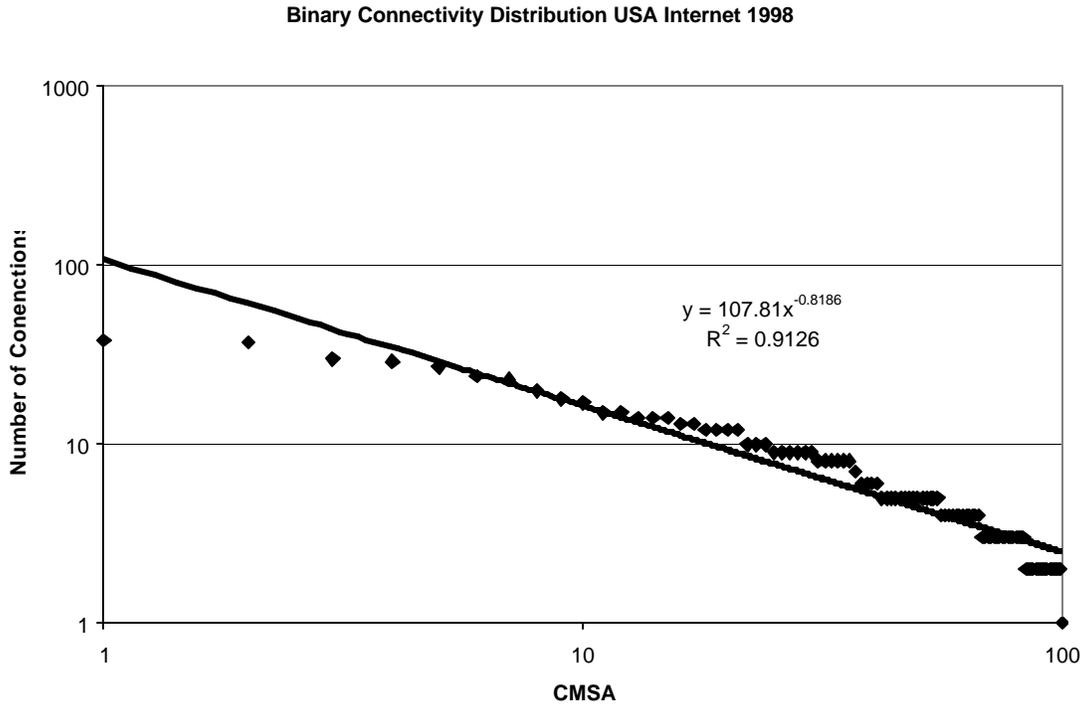

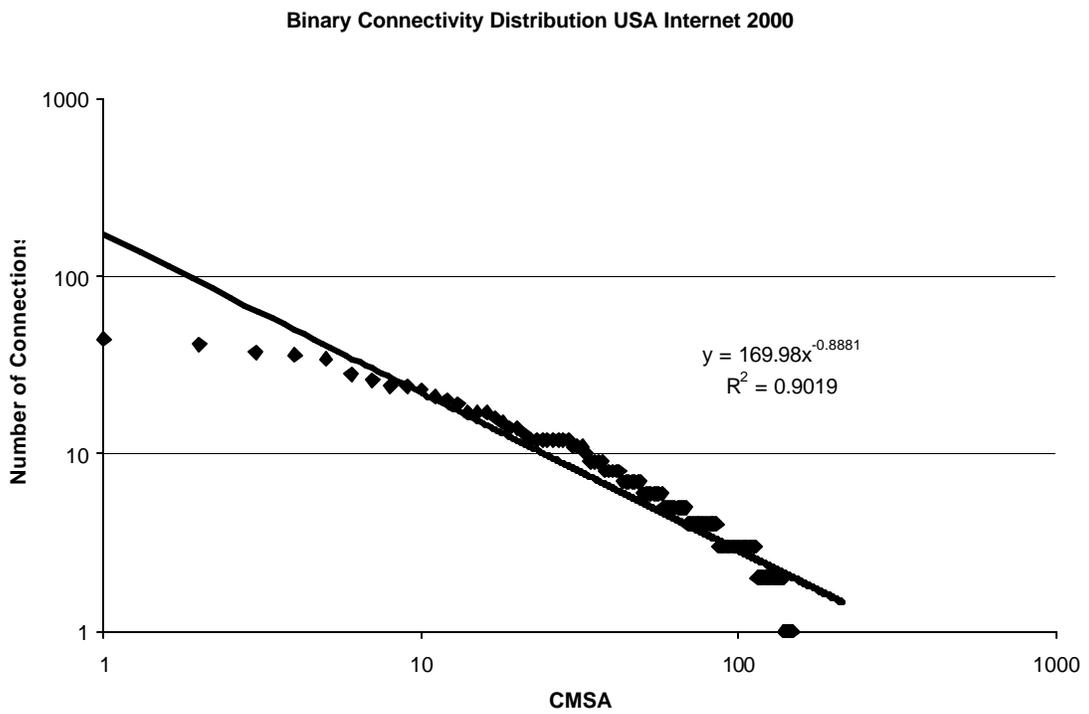

Figure 9. Matrix Construct of the Spatial Small World Model

|    |    | r1 | | | r2 | | | r3 | | |
|----|----|----|----|----|----|----|----|----|----|----|
|    |    | n1 | n2 | n3 | n4 | n5 | n6 | n7 | n8 | n9 |
| r1 | n1 | 0 |   |   |   |   |   |   |   |   |
|    | n2 |   | 0 |   |   |   |   |   |   |   |
|    | n3 |   |   | 0 |   |   |   |   |   |   |
| r2 | n4 |   |   |   | 0 |   |   |   |   |   |
|    | n5 |   |   |   |   | 0 |   |   |   |   |
|    | n6 |   |   |   |   |   | 0 |   |   |   |
| r3 | n7 |   |   |   |   |   |   | 0 |   |   |
|    | n8 |   |   |   |   |   |   |   | 0 |   |
|    | n9 |   |   |   |   |   |   |   |   | 0 |

r = region
n = city
light gray = intra-regional links
dark gray = inter-regional links

Figure 10. Hierarchical Network Clustering

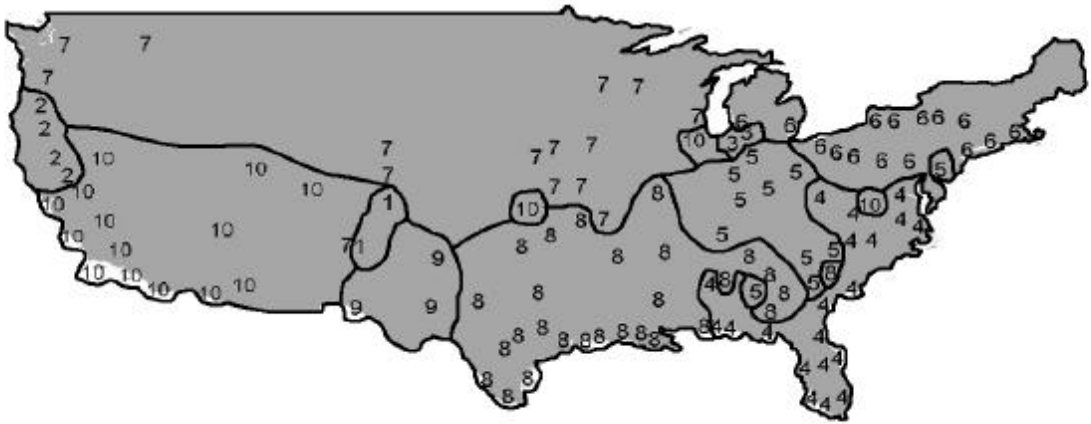

Figure 11. Algorithm's Spatial Output

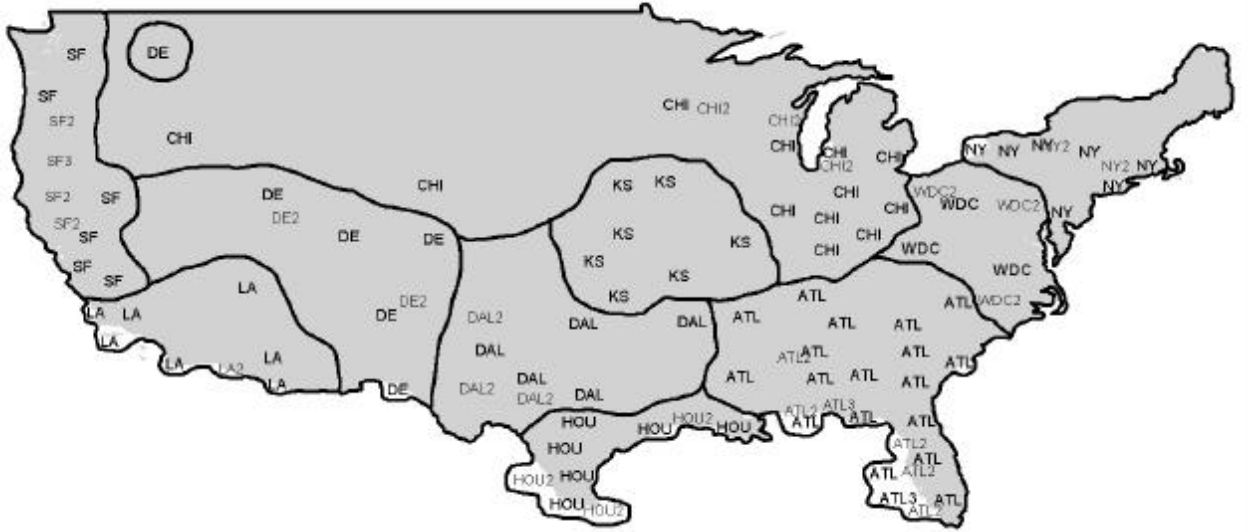